\documentclass[
 aps,
 11pt,
 final,
 notitlepage,
 oneside,
 twocolumn,
 nobibnotes,
 nofootinbib,
 superscriptaddress,
 noshowpacs,
 centertags]
{revtex4}
\usepackage{graphicx}

\def\squareforqed{\hbox{\rlap{$\sqcap$}$\sqcup$}}

\def\sq{\ifmmode\squareforqed\else{\unskip\nobreak\hfil
\penalty50\hskip1em\null\nobreak\hfil\squareforqed
\parfillskip=0pt\finalhyphendemerits=0\endgraf}\fi}

\def\sun{\hbox{$\odot$}}

\def\utw{\smash{\rlap{\lower5pt\hbox{$\sim$}}}}

\def\udtw{\smash{\rlap{\lower6pt\hbox{$\approx$}}}}

\def\diameter{{\ifmmode\mathchoice
{\ooalign{\hfil\hbox{$\displaystyle/$}\hfil\crcr
{\hbox{$\displaystyle\mathchar"20D$}}}}
{\ooalign{\hfil\hbox{$\textstyle/$}\hfil\crcr
{\hbox{$\textstyle\mathchar"20D$}}}}
{\ooalign{\hfil\hbox{$\scriptstyle/$}\hfil\crcr
{\hbox{$\scriptstyle\mathchar"20D$}}}}
{\ooalign{\hfil\hbox{$\scriptscriptstyle/$}\hfil\crcr
{\hbox{$\scriptscriptstyle\mathchar"20D$}}}}
\else{\ooalign{\hfil/\hfil\crcr\mathhexbox20D}}%
\fi}}

\begin{document}

\title{GLOBULAR CLUSTERS LOST by the SAGITTARIUS DWARF SPHEROIDAL GALAXY}

\author{\firstname{N.~R.}~\surname{Arakelyan}}
 \email{n.rubenovna@mail.ru}
 \affiliation{Lebedev Physical Institute, Russian Academy of Sciences, Moscow, 117997  Russia}

\author{\firstname{S.~V.}~\surname{Pilipenko}}
 \affiliation{Lebedev Physical Institute, Russian Academy of Sciences, Moscow, 117997  Russia}
 
\author{\firstname{M.~E.}~\surname{Sharina}}
 \affiliation{Special Astrophysical Observatory, Russian Academy of Sciences, Nizhnij Arkhyz,  369167 Russia}


\begin{abstract}
\textbf{Abstract} -- In this work a search was carried out for globular clusters belonging to the Sagittarius (Sgr) tidal stream using the analysis of spatial positions, radial velocities relative to the Galactic Standard of Rest ($V_{GSR}$),proper motions and ratio of ``age -- metallicity'' ([Fe/H]) for globular clusters and for stars in the tidal stream. As  a  result,  three  categories  of  globular  clusters  were  obtained:  A --most  certainly  in  the  stream:  $Terzan$~8, $Whiting$~1, $Arp$~2, $NGC$~6715, $Terzan$~7, $Pal$~12;  B -- kinematic  outliers:  $Pal$~5, $NGC$~5904, $NGC$~5024,  $NGC$~5053,  $NGC$~5272, $NGC$~288;  C -- lowest  rank  candidates:  $NGC$~6864, $NGC$~5466, $NGC$~5897,   $NGC$~7492, $NGC$~4147.
\end{abstract}

\maketitle
\section{Introduction}  
 It  is  of  great  interest  to  study  the  relationship between  the  evolution  of  galaxies  and  their  environment.  In the standard $\Lambda$CDM cosmological model \citep{1974ApJ...189L..51P}, galaxies are formed gradually by hierarchical merging: first, small -- mass objects are formed, then more massive by merging of them. The Milky Way is no exception.

Some dwarf satellite galaxies being in a long interaction with the Galaxy begin to partially disrupt and merge with the Galaxy. Due to the high velocities of motion, tidal tails arise and gas, dust, stars and globular clusters (GCs) get there. With the close passage of a dwarf galaxy (tidal debris) near the center of the Galaxy due to accretion, the process of transferring its stars and GCs to the Milky Way begins, which become the building material for it.

Near our Galaxy there are a number of tidal streams \citep[ see e.g.][]{1996ApJ...459L..73M,1999Natur.402...53H,2002ApJ...569..245N,2004ApJ...615..732R,2006ApJ...642L.137B,2006ApJ...645L..37G,2006ApJ...641L..37G,2006ApJ...636L..97D, 2007ApJ...658..337B,2008MNRAS.389.1391S,2009ApJ...698..567S,2009ApJ...700L..61N,2010ApJ...711...32N, 2011ApJ...728..102W}  and one of them is the tidal stream of Sgr \citep{1994Natur.370..194I,1995AJ....109.2533D,1997AJ....113..634I,2002AJ....124..915B,2003AJ....125..188B,2003ApJ...599.1082M,2003ApJ...596L.191N,2004AJ....128..245M,2004A&A...414..503B,2007ApJ...667L..57S,2007A&A...466..181C,2009AJ....137.3809C,2010ApJ...718.1128L,2014MNRAS.445.2971C,2017A&A...600A.118C,2017MNRAS.467.1329N,2019ApJ...874..138L}. Sgr dwarf spheroidal galaxy (Sgr dSph) is currently in the process of accretion, so due to its long interaction with the Milky Way some GCs are stripped from it and now lie scattered throughout the halo of the Galaxy.

There are many works in the literature searching for GCs that could have formed in Sgr dSph \citep[e.g.][]{1995AJ....109.2533D,1995MNRAS.275..429L,1999IAUS..192..409I,2000AJ....120.1892D,2001AJ....122.1916D,2002AJ....124..915B,2002ApJ...564..736P,2002ApJ...573L..19M,2003ApJ...596L.191N,2003AJ....125..188B,2004AJ....128..245M,2004MNRAS.355..504M,2004ASPC..327..255M,2004AJ....127.3394F,2007A&A...466..181C,2009AJ....137.3809C,2010MNRAS.404.1203F,2010ApJ...718.1128L,2014MNRAS.445.2971C,2014MNRAS.437..116B,2017A&A...598L...9M,2018ApJ...862...52S,2019A&A...630L...4M,2020MNRAS.tmp..233F,2020A&A...636A.107B}. and in each of them a number of such objects is suggested  (e.g., Pal 2, Pal 5, Pal 12, NGC 2419, NGC 5634, Whiting 1, NGC 4147, NGC 5053, NGC 6715, Arp 2, Terzan 7, Terzan 8, NGC 7492, AM 4, NGC 5824)(see also \cite{2019ARep...63..274M} and references in this article).

In this work, we use our own method to check which GCs belong to the Sgr tidal stream using data on their proper motions, which were not available before and appeared recently in the literature (see \cite{2019MNRAS.482.5138B}) after release of data from Gaia DR2 \cite{2018A&A...616A..10G,2018A&A...616A...1G,2018A&A...616A..12G}. We also use the tidal tail simulation results of Sgr dSph.  There are several such models in the literature \citep[e.g.][]{2010ApJ...714..229L,2017ApJ...836...92D}, but  we  use  data from the \cite{2010ApJ...714..229L} model (hereafter LM10a), since this data can be found in the public domain. In addition, in the LM10a model, the  arms of the tidal stream are in good agreement with observational data, while for the \cite{2017ApJ...836...92D} model there is a discrepancy for the leading arm. The model presented in the LM10a includes the potential  of the Galaxy, which consists of three components: disk, spherical and triaxial halo. The last component is elongated in a direction almost perpendicular to the Galactic disk. For the Sagittarius dwarf spheroidal galaxy in LM10a, the galactic coordinates are $(l,b)=(5.6^\circ , -14.2^\circ )$ and the distance from the Sun is 28~kpc \cite{2007ApJ...667L..57S}. The satellite's orbit at the time of observation is characterized by the direction of the pole $(l_p,b_p)=(273.8^\circ , -14.5^\circ )$ , radial velocity of 171~km s$^{-1}$ . The remaining parameters of the satellite were found by fitting numerical models of the tidal destruction of the   satellite   to   the   data   on   the   velocities   of   the observed stars of the trailing arm of stream, followed by testing the coincidence of the model results and the observed positions of the stars also for the leading arm. The data for the observed stars of the stream were obtained from the 2MASS and SDSS surveys \cite{2003ApJ...599.1082M,2004PASA...21..197M,2004ASPC..327..239L,2005ApJ...619..807L,2006ApJ...642L.137B,2009ApJ...700.1282Y}. So, the found three--dimensional velosity of the Sgr dSph is $(V_x,\, V_y,\, V_z) = (230,\, -35,\, 195)$ km s$^{-1}$, the current mass is $2.5^{+1.3}_{-1.0}\times 10^8$ M$_\odot$. Model LM10a is an improved version of past models (see \cite{2004ASPC..327..239L,2005ApJ...619..807L}), since it is best approximated to the observational data of the tidal stream. For example, in the previous tidal disruption model Sgr dSph, the authors presented three different models for the Galactic halo (axisymmetric in the disk plane; oblate; elongated in the direction perpendicular to the disk), but none of them could simultaneously reproduce angular positions, distances and radial velocities. In the model LM10a, they modeled a non-axisymmetric, almost flattened halo, which reproduced all the limitations within reasonable accuracy.
Using it together with real data on class M giants (see  \cite{2004AJ....128..245M,2004ASPC..327..239L, 2005ApJ...619..807L} \footnote{\url{http://faculty.virginia.edu/srm4n/Sgr/index.html}}) for the stars of the stream, we get more opportunities to search for GCs belonging to Sgr dSph.

This paper is organized as follows: In Section~2 we describe  the  method  of  GCs  selection  using  nearest neighbors, we apply it to identify GCs in the arms of the tidal stream of Sgr, check the membership to the stream by radial velocities, by proper motions and bythe distribution of ``age -- metallicity''; we present discussion and conclusions in Section~3.

\section{Identifying the GCs Belonging to the Tidal Stream}\label{sec:2} 
\subsection{Spatial Positions}\label{sec:2.1} 
\begin{figure*}
 \includegraphics[width=0.325\textwidth]{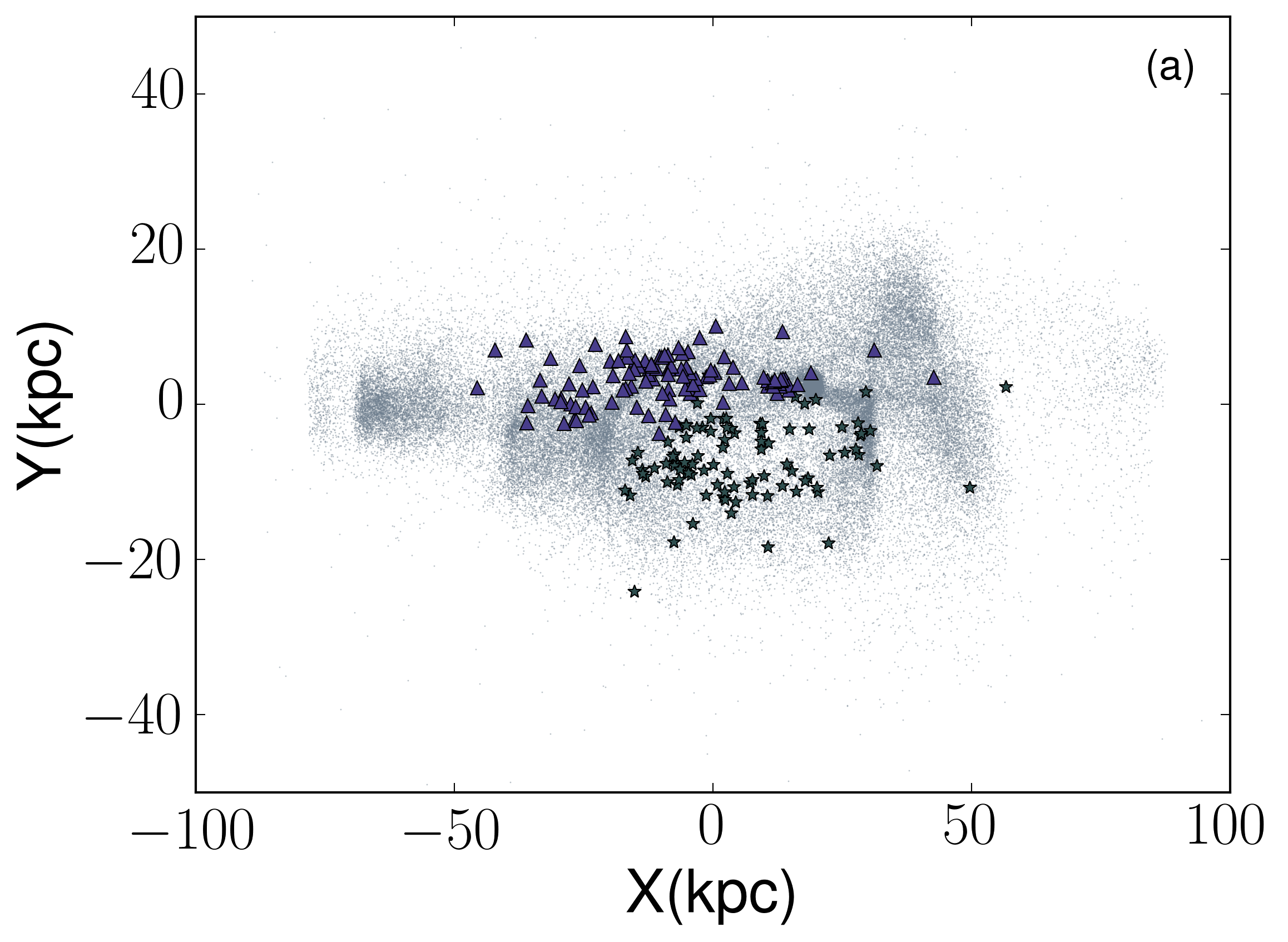} 
 \includegraphics[width=0.325\textwidth]{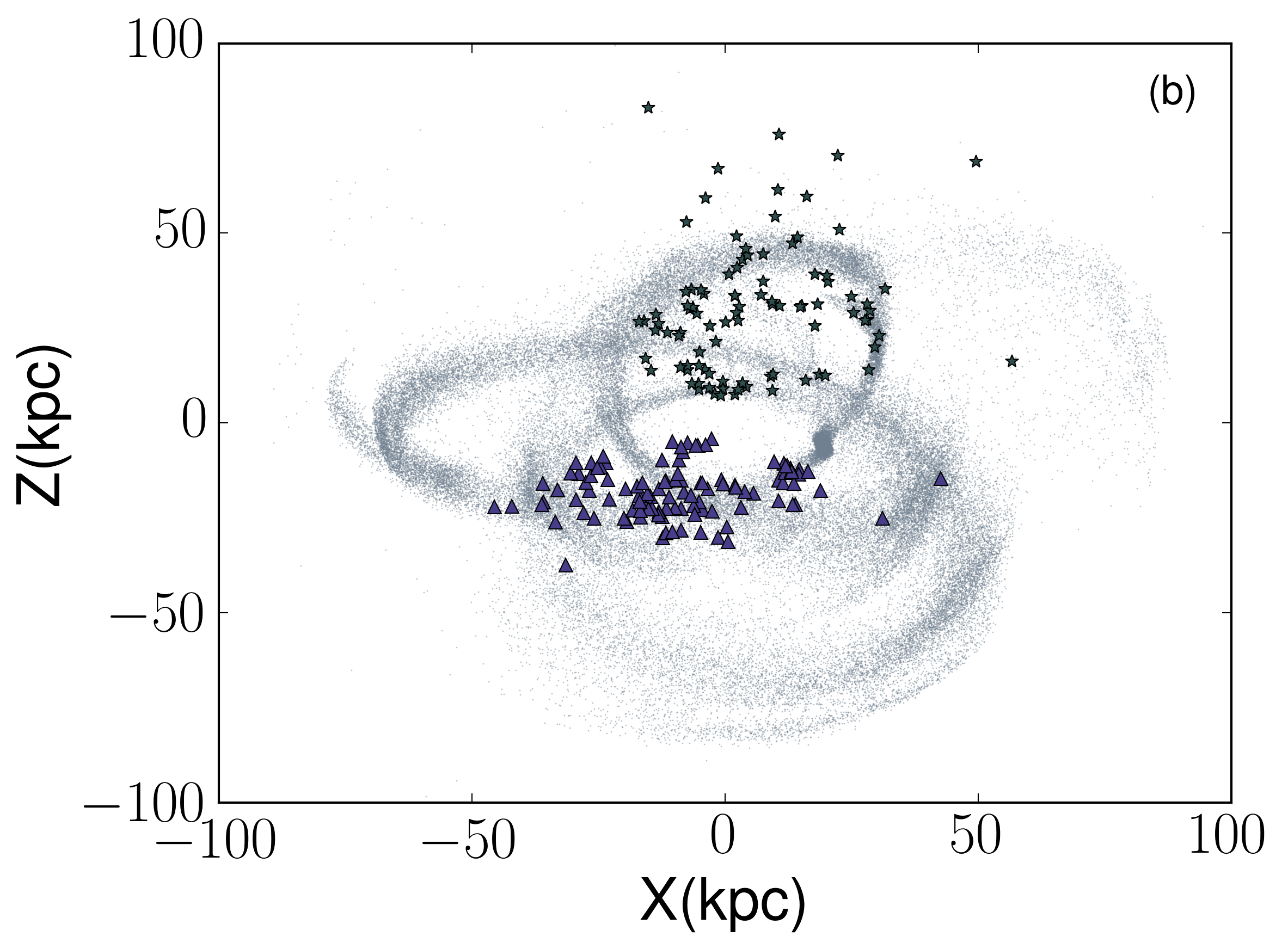} 
 \includegraphics[width=0.325\textwidth]{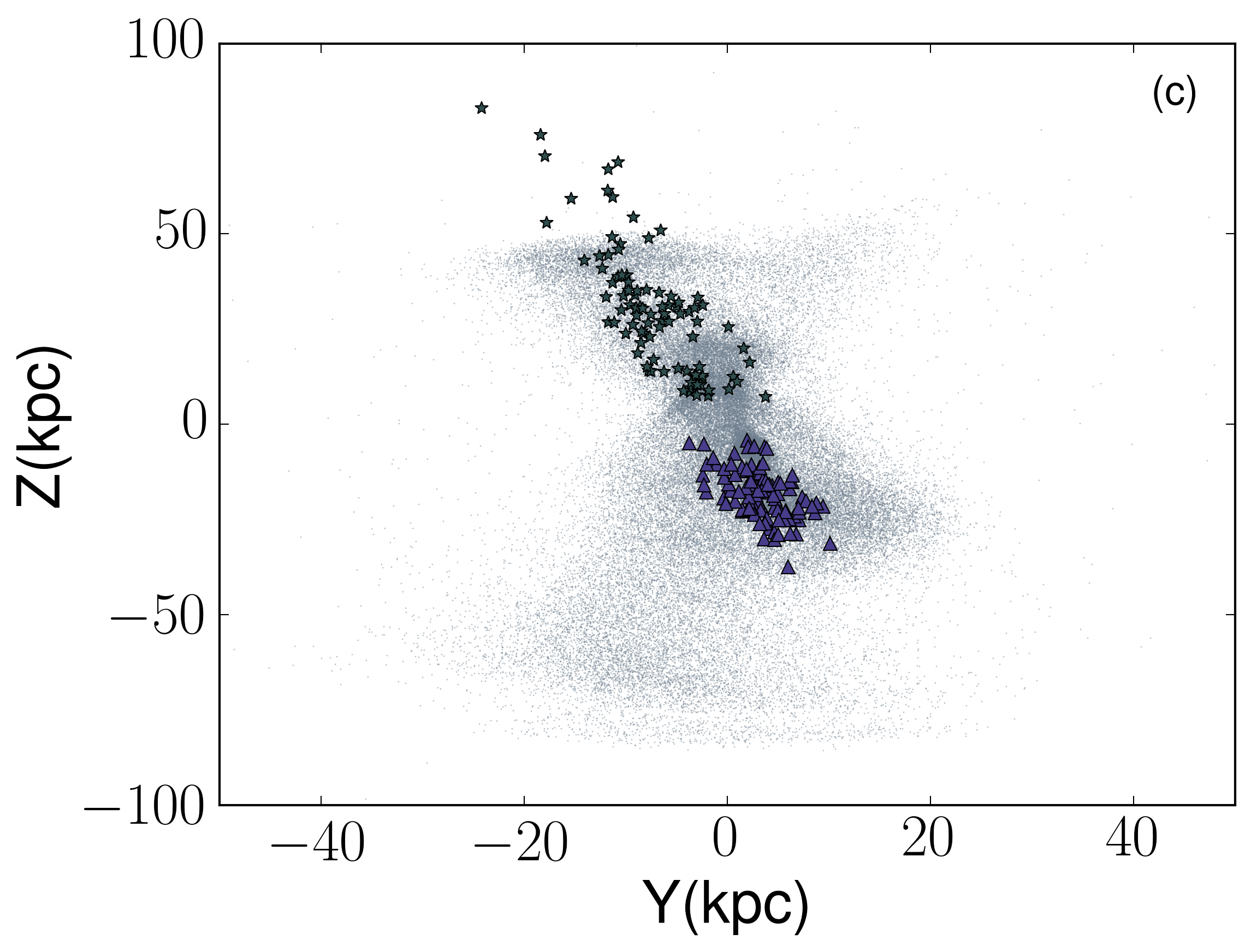} 
 \caption{ Sgr tidal stream in three projections. The gray dots represent the  Sgr  stream model (LM10a), star marks show stars in the leading arm and triangles show stars in the trailing arm (observed data). XYZ -- Cartesian coordinates relative to the center of the Galaxy. Z is aligned with the Galactic pole and coordinates of the Sun are -8.34, 0, 0 kpc \citep{2014ApJ...783..130R}}
 \label{fig:1}
\end{figure*}

The  simulation  results  from  article  LM10a  show that as a result of the destruction of the original object,the  stars  which  left  it  form  a  stream  of  characteristic shape
 (Fig.~\ref{fig:1}). The same fate should befall the clusters that previously belonged to the satellite galaxy. Most of the stars are concentrated in a relatively narrow stream, while the rest are distributed quite widely throughout the Galaxy. In other words, with a certain non -- zero probability, a star or a GC from the Sgr dSph can appear both in a narrow stream and outside it. Therefore, to search for GCs that previously belonged to the Sgr dSph, we use information on the spatial density of stars from the LM10a model.

To find out if a cluster can belong to the stream, we measure the density of stars from the stream around each cluster and compare it with the density of the stars of the stream around each stream star. Since the density of stream stars varies greatly in space, we do not measure the density in a sphere of a fixed radius, but instead fix the number of stars in a sphere equal to $n$. In other words, the radius of a sphere is the distance to the $n$th nearest neighbor. We took $n=6$, but our results are weakly dependent on $n$. The density itself is not necessary to calculate, we use the distance $d_6$ to the sixth neighbor for further analysis.

The distribution of distances to the sixth neighbor gives us the probability for a star in the LM10a model to have such a neighbor at a certain distance or less. The probability for the distance $d_6$ is equal to the ratio of the number of particles in the model having a sixth neighbor at a distance less than $d_6$ to the total number of particles. For all 157 GCs, we find the distance to the nearest sixth neighbor -- star from the LM10a model, and determine  the  probability  corresponding to  this  distance  from  the  distribution  of  distances between  the  stars  in  the  model,  as  indicated  above.
These probabilities ranged from $10^{-4}$ to $0.6$. However, one should not assume that the probability found is the probability for the GC to belong to the stream. In the model stream itself, there are stars with a probability of up to $10^{-5}$, which, nevertheless, belong to the stream by model construction. If there are N objects that belong to the stream, then the probability determined through the $d_6$ distribution for them can be of the order of $ 1/N $ or higher. Also, the distribution of stars in the LM10a model may slightly differ from the real stream due to differences of the model and real gravitational potentials of the Galaxy. Therefore, in what follows we consider all GCs with a probability of more than 0.01 as candidates for belonging to the straem. As a result, we get 17 GCs candidates with probabilities from 0.013 to 0.586 (see. Table~\ref{tab:1}).

The tidal stream from the Sgr dSph can be divided into two arms: leading and trailing. In order to determine which of these arms each of the 17 clusters belongs to, we find distances $d_6$ for each star in the arms of tidal stream (stars from the catalog   http://faculty.virginia.edu/srm4n/Sgr/ index.html). After that we take the median of these distances and check which GCs are at this or smaller distance from the stars in the stream. In other words, we determine the belonging to the arm based on data on the nearest observable stars from the stream (Both arms are also present in the LM10a model, but the star data is given in one list and there is no division of stars by arms).

The number of stars in the leading and  trailing tidal arms is 94 and 108, respectively. As a result of our analysis of the spatial distribution of the GC, for the observed stars in the stream, we get eight GCs in the leading and eight GCs in the trailing arms of the tidal stream of Sgr. Only one GC out of 17 was not found in any of the arms and this cluster is $NGC$~6715. This is no coincidence, since $NGC$~6715 does not belong to the arms and is located directly in the center of the Sgr dSph \cite{2010ApJ...718.1128L,2019A&A...630L...4M}. The coordinates of Sgr dSph itself \citep{2010ApJ...714..229L} and $NGC$~6715 \citep[][2010 edition]{2013ApJ...772...82H,1996AJ....112.1487H}\footnote{\url{http://physwww.mcmaster.ca/~harris/Databases.html}} are very close: for Sgr $L_{Sgr} = 5.6^\circ$, $B_{Sgr} = -14.2^\circ$, $d_{\odot Sgr} = 28$~kpc, $R_{Sgr}\approx 20$~kpc, while for $NGC$~6715 $L = 5.61^\circ$, $B = -14.09^\circ $, $d_\odot = 26.5$~kpc, $R \approx 19$~kpc. 

\subsection{Kinematics of GCs}\label{sec:2.2}
After obtaining the list of 17 GCs, we checked  how well the radial velocities of the GCs coincide with the velocities of the stars in observed and model catalogues. Radial velocities for stars (for observable data and for model data) were taken from \cite{2007A&A...464..201M,2010ApJ...714..229L} (http://faculty.virginia.edu/srm4n/Sgr/index.html). For GCs, radial velocities were taken from \cite{2019MNRAS.482.5138B}. All the velocities in our paper are given in the Galactic Standard of Rest, i.e. they are corrected for the Galactic rotation in the Solar neighborhood. The radial velocity in the Galactic Standard of Rest $V_{GSR}$ is connected to the velocity in the Local Standard of Rest $V_{LSR}$ with the formula:

\begin{equation}
 V_{GSR} = V_{LSR} + 220 \sin (l) \cos(b)\ \text{km/s}. 
\end{equation}

\begin{figure*}
 \includegraphics[width=0.49\textwidth]{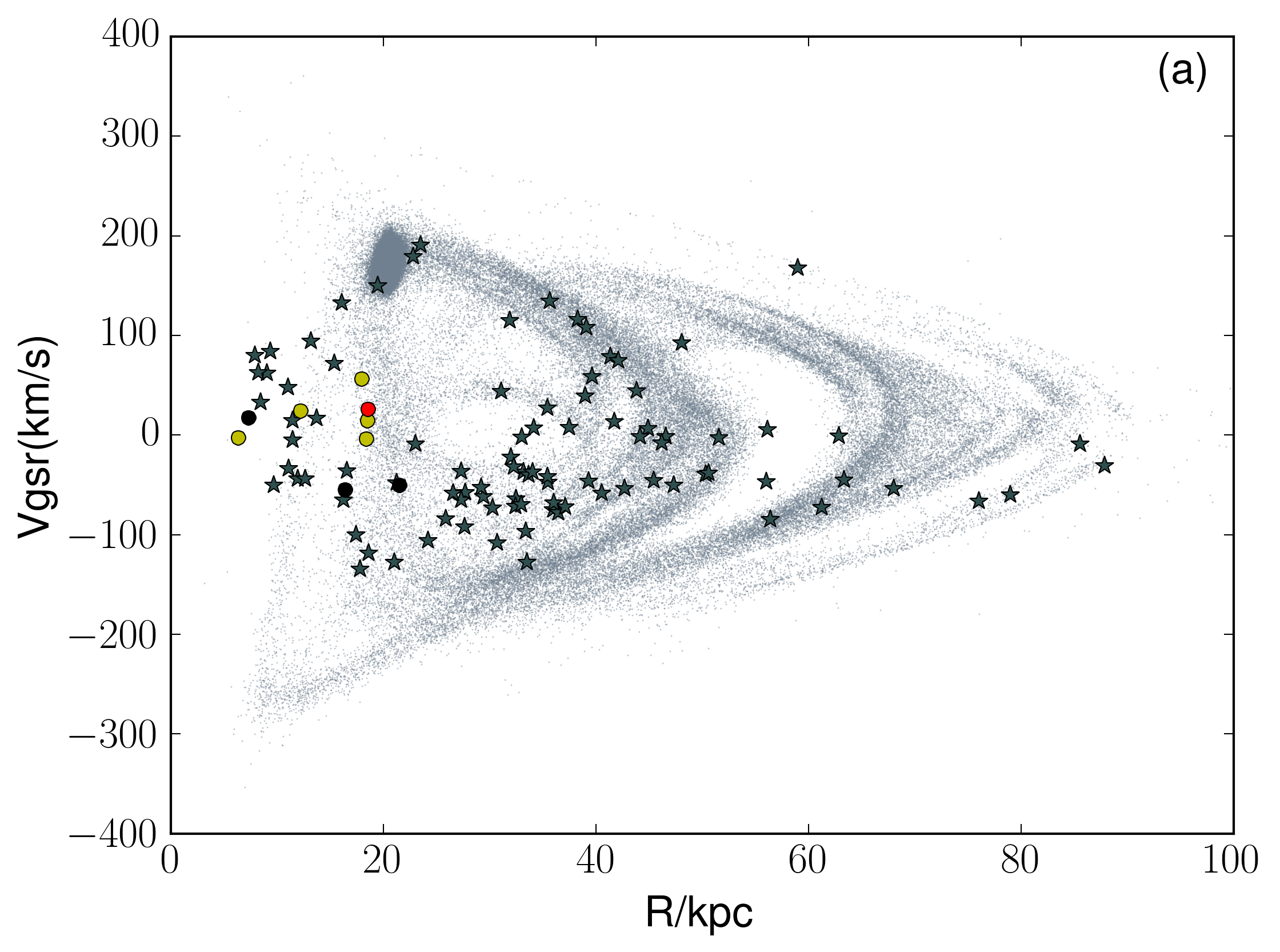} 
 \includegraphics[width=0.49\textwidth]{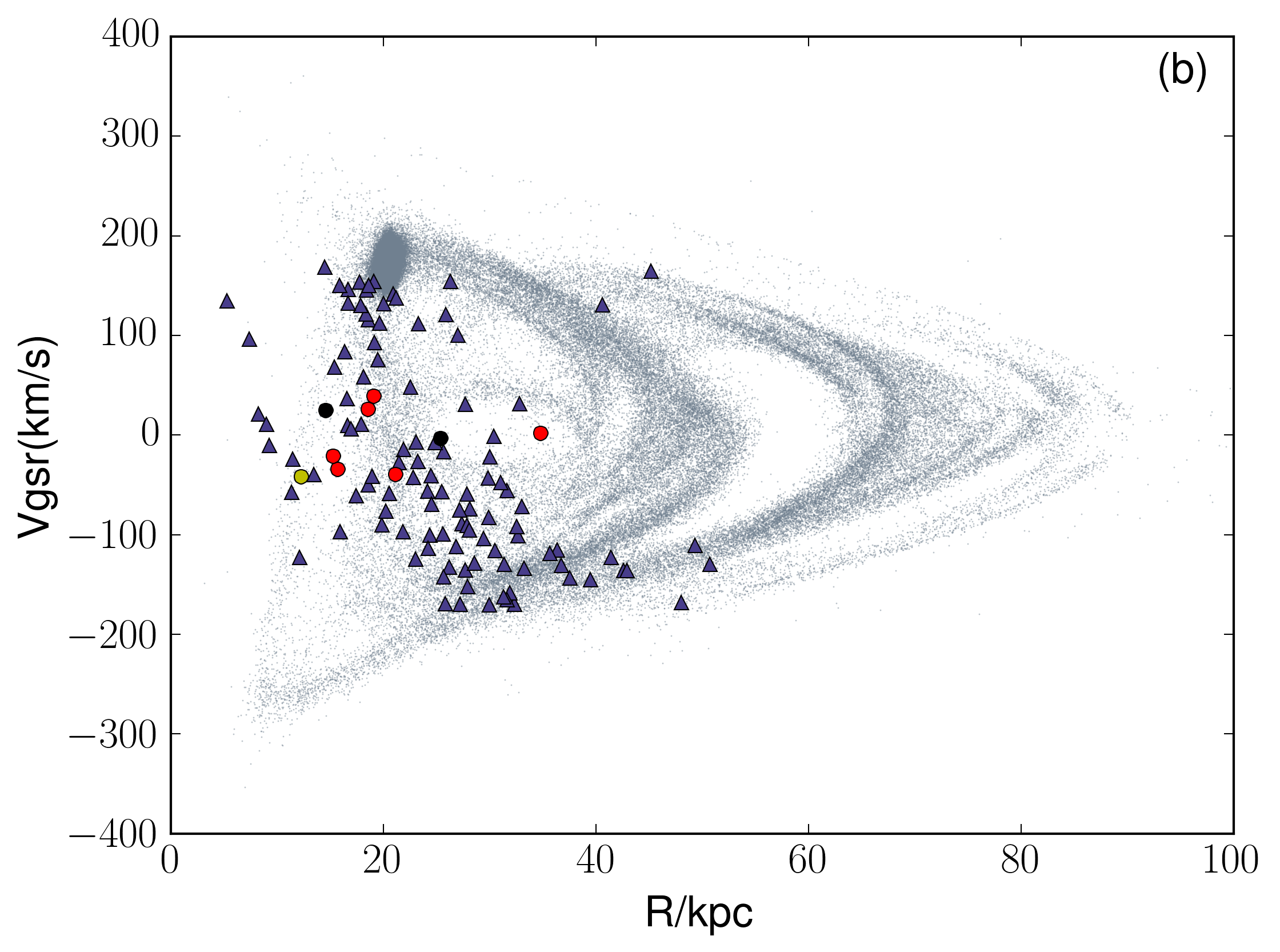} 
 \caption{ Radial velocities depending on the Galactocentric distance for the GCs of our sample (red, yellow and black dots), the observed stars from the leading (stars, panel (a)) and trailing arms  (triangles, panel (b)),  respectively, and the Sgr tidal stream model LM10a ( gray dots). The red dots show GCs that most likely belong to the stream  (five GCs in trailing arm, one more GC, which is shown in both figures ((a), (b)) ($NGC$ 6715) and which is directly next to the $Sgr$ $dSph$). The yellow dots are the GCs candidates -- members of the stream ( five GCs in the leading arm and one GC in the trailing arm). Black dots show the remaining five GCs.}
 \label{fig:2}
\end{figure*}

The results are shown in Figure~\ref{fig:2} and  Table~\ref{tab:1}.  Figure~\ref{fig:2} shows the distribution of GCs and stars from the stream by radial velocities.

\begin{table*}[]
 \caption{ The probability that GCs belong to the Sgr stream, radial velocities for 17 GCs ($ V_ {GSR} $) in the stream and average radial velocities for their nearest six stars in the stream  ($ V_ {GSR} ^ * $ for stars from model LM10a and $ V_ {GSR} ^ {**} $ for observed stars). Errors are three standard deviations. In addition, indicated the arm of the tidal stream to which the cluster belongs and the type of GCs according to the classification \citep{2005MNRAS.360..631M} }
 \label{tab:1} 
 \medskip
 \begin{tabular}{|l|c|c|c|c|c|c|}
  \hline
  \multicolumn{1}{|l|}{Name} & Probability & $V_{GSR}$  & $\langle V_{GSR}^*\rangle$ & $\langle V_{GSR}^{**}\rangle$ & Arm & Type \\
 & & (km s$^{-1}$) & (km s$^{-1}$) & (km s$^{-1}$) & &\\
  \hline
NGC 6864 (M75)& 0.013 & $-189.08$ & 129.40$\pm$377.31& 147.12$\pm$99.31   & Trailing & OH \\
NGC 5466 & 0.016 & 106.93  &  $-32.85$\,$\pm$102.40 & $-53.47$\,$\pm$189.40 & Leading  & YH \\ 
NGC 288  & 0.019 & $-44.83$ & 154.59$\pm$260.99 & $-34.31$\,$\pm$167.99 & Trailing & OH \\
NGC 5272 (M3) & 0.036 & $-147.28$ & $-72.07$\,$\pm$213.21 & $-36.70$\,$\pm$135.21 & Leading  & YH \\
NGC 5053 & 0.048 & 42.77 & $-35.98$\,$\pm$274.16  & $-51.16$\,$\pm$193.16 & Leading  & YH \\
NGC 5897 & 0.052 & 101.31 & $-264.20$\,$\pm$126.87 & 53.56$\pm$132.87 & Leading & OH \\
NGC 5024 (M53) & 0.060 & $-62.85$ & $-4.01$\,$\pm$102.88 & $-63.52$\,$\pm$210.88 &  Leading & OH \\
NGC 7492  & 0.071 & $-176.70$ & 156.99$\pm$264.21 & $-5.92$\,$\pm$115.21  & Trailing & OH \\
Pal 12  & 0.076 & 27.91 & 104.26$\pm$110.60 & 69.52$\pm$116.60  & Trailing & SG \\
NGC 5904 (M5) & 0.079 & 53.70 & $-189.19$\,$\pm$471.07 & 25.57$\pm$207.07 & Leading & OH \\
Pal 5 & 0.083 & $-58.60$ & $-124.22$\,$\pm$159.52 & 74.13$\pm$174.52 & Leading & YH \\
Terzan 7 & 0.092 & 159.45   & 183.57$\pm$172.53 & 151.15$\pm$94.53   & Trailing & SG \\
NGC 4147 & 0.100 & 179.52 & 51.25$\pm$89.13 & $-91.64$\,$\pm$149.13 & Leading & SG  \\
NGC 6715 (M54) & 0.144& 143.06 & 187.46$\pm$91.94 &  & & SG  \\  
Arp 2 & 0.252 & 123.01 & 169.73$\pm$93.48 & 146.45$\pm$87.48 & Trailing & SG \\
Whiting 1 & 0.275 & $-130.41$ & $-114.02$\,$\pm$118.97 & $-106.11$\,$\pm$139.97 & Trailing & UN \\
Terzan 8 & 0.586 & 148.53 & 164.84$\pm$87.39 & 146.45$\pm$84.39 &  Trailing & SG \\                       
  \hline   
  \end{tabular}
\end{table*}

Table~\ref{tab:1} shows the probabilities for each of the 17 cluster candidates (ordered by increasing probability) belonging to the stream; the radial velocities relative to the Galactic Standard of Rest ($V_{GSR}$) for 17 GCs; for each GC average radial velocities for  nearest six stars in the stream: for stars from model LM10a ($V_{GSR}^*$)   and for stars from real data $V_{GSR}^{**}$.  There is also shown in which arm these clusters were discovered and to which type according to the classification \citep{2005MNRAS.360..631M} they belong. After examining Table~\ref{tab:1}, we can find out for which GCs the radial velocities are close to those of the stars of the stream. Nearest  observed stars are too far from $NGC$~6715. Therefore, the table does not provide data on the average radial velocity of the stars closest to this cluster.

Errors for radial velocities are composed of three components: (1) for each cluster, we took three standard deviations from the average radial velocities for the nearest six stars in the model stream, (2) three standard deviations from the results of measurements of GCs velocities \cite{2019MNRAS.482.5138B} and (3) additional error introduced to account for the discrepancy between the real and model potentials of the Galaxy. The latter was estimated by comparing the radial velocities of the observed stars and the nearest stars to them from the model.

Table~\ref{tab:1} shows that  for 13 GCs, the radial velocities within the errors coincide with the velocities of the stars from the model. For the observed stars, this is true for 13 GCs. For 11 GCs, the radial velocities coincide with the data of both the model and the catalog of observed stars. We will add $ NGC $ ~ 6715 to this list, since its radial velocity coincides with the velocities of the nearest particles from the model, and we cannot verify the coincidence of the observed stars, because, as mentioned above, this cluster does not belong to the arms and is located next to Sgr dSph. As a result, we get 12 GCs (($NGC$~288; $NGC$~5272; $NGC$~5053; $NGC$~5024; $Pal$~12; $NGC$~5904; $Pal$~5; $Terzan$~7; $NGC$~6715; $Arp$~2; $Whiting$~1 and $Terzan$~8)), which according to radial velocities belong to the tidal stream of Sgr.

More information about the GC and their belonging to the Sgr stream can be obtained from 3D velocities of the GCs and stars in the stream. Table~\ref{tab:2} shows the spatial velocities $V_x$, $V_y$, $V_z$ for 17 GCs \citep{2019MNRAS.482.5138B}, where X axis points from the Galactic centre towards the Sun, Y points in the direction of Galactic rotation at the  solar position, and Z points towards  the north Galactic pole. We used data from \citep{2019MNRAS.482.5138B} in our work, however we compared them with data from \cite{2019MNRAS.484.2832V} and found that they are in good agreement with each other.

\begin{figure*}
 \includegraphics[scale=0.43]{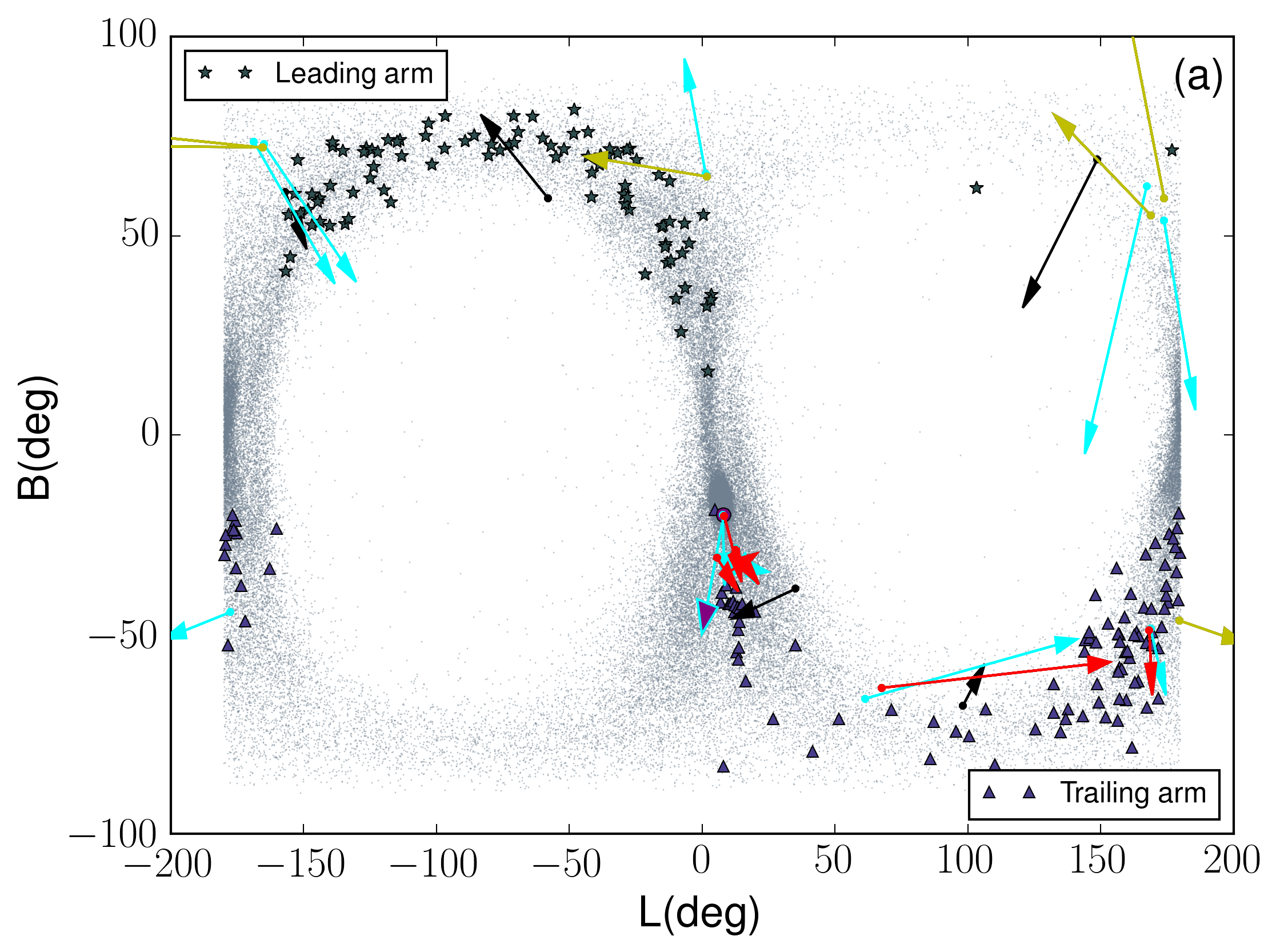}
 \includegraphics[scale=0.43]{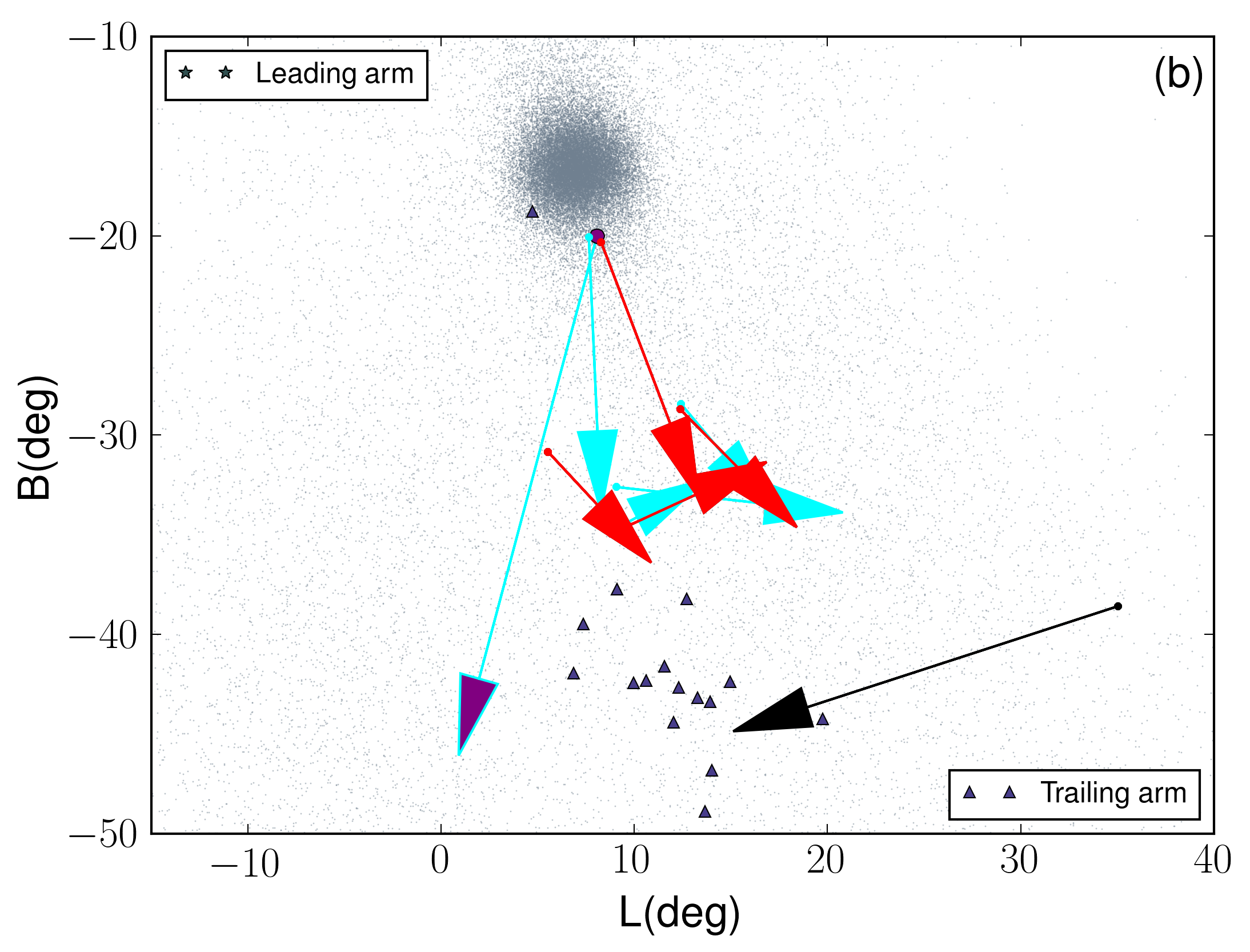}
 \caption{ Map of the sky in Galactocentric Galactic coordinates. Coordinates of the Sun are -8.34, 0, 0 kpc \citep{2014ApJ...783..130R} (a). Enlarged portion of the sky map near Sgr dSph (b). Gray dots represent the Sgr tidal stream model (LM10a). Star marks show observed stars in the leading arm and triangles in the trailing arm. The red, yellow, and black arrows show proper motions for six GCs (category A) belonging to the stream, for six candidates (category B) and for the remaining five GCs (category C), respectively. The cyan arrows correspond the average proper motion of the six nearest model stars in the stream and cyan arrow with a purple tip is proper motion of Sgr dSph. }
 \label{fig:3}
\end{figure*}

Similarly to what was done above for radial velocities, we found the average  Galactocentric velocities $<V_x> $, $ <V_y> $, $ <V_z> $  for the six nearest neighboring particles of the model for each of the 17 GCs. The table also presents errors (three standard deviations, $ 3\sigma_x $, $ 3\sigma_y $, $ 3\sigma_z $), which were calculated in the same way as for radial velocities. The proper motions of the GCs with respect to the motions of the nearest stars in the stream are presented in Figure~\ref{fig:3}, which presents a map of the sky in Galactocentric coordinates L, B. Using the data L, B for 17 GCs from Table~\ref{tab:3}, it is possible to determine the location of each GC on the map of the sky (see Fig.~\ref{fig:3}). From Table~\ref{tab:2} we see that the spatial velocities of the GCs $ Pal$~12; $ Terzan $~7; $ NGC $~6715; $ Arp $~2; $ Whiting $~1 and $ Terzan $~8 are in good agreement with the average proper motions of their nearest neighbours and we can conclude that these six clusters, by their 3D velocities, belong to the Sgr stream.

If we compare the list of  12 GCs obtained by coincidence of radial velocities and a list of six GC with coinciding 3D velocities, we note that six clusters are in both lists and this fact leads to the conclusion that, according to the kinematics, these six clusters ($Pal$~12; $Terzan$~7; $NGC$~6715; $Arp$~2; $Whiting$~1 and $Terzan$~8) exactly belong to the stream (hereafter category A), while the remaining six clusters ($NGC$~288; $NGC$~5272; $NGC$~5053; $NGC$~5024; $NGC$~5904  and $Pal$~5) for which there is a discrepancy in radial velocities or in their proper motions, become candidates (hereafter category B). The remaining five GCs out of 17 ( $NGC$~6864; $NGC$~5466; $NGC$~5897; $NGC$~7492 and $NGC$~4147), which do not belong to the  stream by kinematic, become less probable candidates (hereafter category C).

\begin{table*}
 \caption{Proper motions for 17 GCs \citep{2019MNRAS.482.5138B} in the stream and average proper motions for their nearest six stars in the model stream}
 \label{tab:2} 
 \medskip
 \begin{tabular}{|l|c|c|c|c|c|c|}
  \hline
  \multicolumn{1}{|l|}{Name} &  $V_x$  &  $V_y$ & $V_z$ & $\langle V_x \rangle$ & $\langle V_y \rangle$ & $\langle V_z \rangle$  \\
    & & &  &  $3\sigma_x$ &   $3\sigma_y$  &   $3\sigma_z$      \\
   & (km s$^{-1}$) & (km s$^{-1}$) & (km s$^{-1}$) & (km s$^{-1}$) & (km s$^{-1}$) & (km s$^{-1}$)  \\
  \hline 
NGC 6864 & 66.92 & $-83.26$ & 49.40 & $-202.82$ & 39.98 & 108.39 \\
   &&&& 478.13 & 113.00 & 221.00 \\
NGC 5466 & 235.28 & $-50.46$ & 232.32 & 341.41 & 10.72 & 62.21 \\
    &&&& 78.38 & 137.24 & 95.24 \\
NGC 288 & 9.89 & $-80.69$ & 50.55 & $-232.32$ & 80.45 & $-176.78$ \\
   &&&& 291.17 & 74.39 & 207.83 \\
NGC 5272 & $-60.46$ & 135.49 & $-134.57$ & 336.64 & $-76.25$ & $-12.86$ \\
   &&&& 108.89 & 258.14 & 204.53 \\
NGC 5053 & $-52.38$ & 148.20 & 35.11 & 333.11 & $-12.02$ & 22.92 \\
   &&&& 68.72 & 90.50 & 281.93 \\
NGC 5897 & 34.13 & $-133.63$ & 88.40 & 378.21 & $-77.91$ & 36.30 \\
    &&&& 99.05 & 142.43 & 160.49 \\
NGC 5024 & $-58.06$ & 158.52 & $-71.86$ & 334.13 & $-8.40$ &  49.14 \\
 & & & & 68.24 & 80.15 & 82.31 \\
NGC 7492  & $-5.14$ & $-95.43$ & 63.68 & $-239.03$ & 62.02 & $-74.85$ \\
 & & & & 135.29 & 171.92 & 257.15 \\
Pal 12 & $-339.19$ & 12.40 & 116.00 & $-328.84$ & $-30.82$ & 105.78 \\
   &&&& 77.72 & 87.35 & 95.45 \\
NGC 5904 & $-304.75$ & 86.82 & $-183.79$ & 284.51 & $-23.71$ & $-7.75$ \\
     &&&& 784.37 & 384.20 & 135.14 \\
Pal 5 & 50.24 & $-170.32$ & $-8.79$ & 303.09  & $-19.26$ & 117.33  \\
   & & & & 94.01 & 124.97 & 231.80 \\ 
Terzan 7    & $-260.69$ & $-5.65$ & 182.9 & $-266.06$ & 16.76 & 175.38 \\
   &&&& 183.56 & 210.95 & 191.66 \\
NGC 4147 & 40.98 & $-10.08$ & 130.87 & 319.89 & $-34.38$ & 30.07 \\
  & & & & 75.53 & 106.73 & 94.40 \\
NGC 6715 & $-229.62$ & 3.26 & 189.18 & $-247.83$ & $-28.67$ & 205.01 \\    
  & & & & 84.02 & 87.77 & 81.35 \\    
Arp 2 & $-251.65$ & $-20.12$ & 180.69 & $-253.95$  & $-31.41$ & 173.64 \\
   & & & & 91.07 & 82.49 & 78.89 \\  
Whiting 1 & $-210.54$ & 36.01 & 12.03 & $-254.12$ & 7.99 & $-5.52$ \\
 & & & & 103.82 & 110.99 & 99.65 \\  
Terzan 8 & $-271.58$ & $-1.88$  & 161.22 & $-265.75$ & $-28.79$ & 176.13 \\
 & & & & 81.59 & 79.37 & 82.73 \\    
  \hline   
  \end{tabular}
\end{table*}

\subsection{Properties of stellar population of Sgr dSph}\label{sec:2.3}
Sgr dSph galaxy has experienced several star forming events  \cite{2007ApJ...667L..57S,2010ApJ...714..229L,2017A&A...605A..46M}.  Based on deep stellar photometry in images taken with the Hubble Space Telescope,  \cite{2007ApJ...667L..57S} concluded that there were at least four stellar populations in Sgr dSph:

 1) $[Fe/H]=-1.8$, $[\alpha/Fe]=+0.2$ with age  about 13~Gyr;

 2) $[Fe/H]=-0.6$, $[\alpha/Fe]=-0.2$ with ages from 4 to 6~Gyr;

 3) $[Fe/H]=-0.1$, $[\alpha/Fe]=-0.2$ with age about 2.3~Gyr;

 4) $[Fe/H]=+0.6$, $[\alpha/Fe]=0.0$ with ages from 0.1 to 0.8~Gyr.

 Here $[\alpha/Fe]$ is the content of $ \alpha $ -- elements \footnote{We use the standard definition, $[X/Y]=log(X/Y)-log(X_{\sun}/Y_{\sun})$,\ where\ X\ and\ Y\ are\ masses\ of\ specific\ elements.}. Later spectroscopic studies of approximately 6600 stars in the center of Srg dSph \cite{2019ApJ...886...57A} proved the existence of three stellar populations: (1) young,  metal--rich $[Fe/H]=-0.04$ and age 2.2~Gyr; (2) middle age, metal--rich $[Fe/H]=-0.29$, 4.3~Gyr and (3) old,  metal--poor $[Fe/H]=-1.41$, 12.2~Gyr.  The  youngest and highest metallicity stars of Sgr dSph are concentrated toward the center of this galaxy. In the center the old and low metallicity GC 
$NGC$~6715 is also located. 

\begin{table*}
 \caption{Galactocentric coordinates \citep{1996AJ....112.1487H} (updated as in 2010 ),  mean age and mean metallicity  (mean value forsamples from the \cite{1996AJ....112.1487H} (updated as in 2010 ); \cite{2010MNRAS.404.1203F,2010ApJ...708..698D,2011ApJ...738...74D,2013ApJ...775..134V}) for 17 GCs. }
 \label{tab:3} 
 \medskip
 \begin{tabular}{|l|c|c|c|c|}
  \hline
  \multicolumn{1}{|l|}{Name} &  L  & B & $\overline{\strut Age}$ & $\overline{\strut [Fe/H]}$    \\
  & ($^\circ$) & ($^\circ$) & ($Gyr$) &  \\
  \hline 
NGC 6864 & 35.04 & $-38.59$ & 9.98$\pm$0.51 & $-1.16$\,$\pm$0.13 \\ 
NGC 5466 & 148.70 & 69.17 & 13.02$\pm$0.48 & $-2.15$\,$\pm$0.12 \\
NGC 288  & 179.69 & $-46.57$ & 11.54$\pm$0.43 & $-1.30$\,$\pm$0.99 \\
NGC 5272 & 168.93 & 55.05 & 11.88$\pm$0.42 & $-1.49$\,$\pm$0.09 \\
NGC 5053 & $-165.48$ & 72.23 & 12.68$\pm$0.47 & $-2.24$\,$\pm$0.16 \\
NGC 5897 & $-57.99$ & 59.37 & 12.30$\pm$1.20 & $-1.82$\,$\pm$0.09 \\
NGC 5024 & $-165.29$ & 72.09 & 12.72$\pm$0.44 & $-2.01$ \,$\pm$0.09 \\
NGC 7492 & 98.06 & $-67.97$ & 12.0$\pm$1.40 & $-1.6$ \,$\pm$0.19 \\
Pal 12   & 67.57 & $-63.43$ & 9.11$\pm$0.57 & $-0.82$\,$\pm$0.02 \\
NGC 5904 & 173.87 & 59.38 & 11.46$\pm$0.44 & $-1.26$\,$\pm$0.08 \\
Pal 5 & 1.76 & 64.84 & 10.9 $\pm$0.86 & $-1.35$\,$\pm$0.08 \\
Terzan 7 & 5.54 & $-30.84$ & 7.65$\pm$0.45 & $-0.49$\,$\pm$0.12 \\
NGC 4147 & $-156.93$ & 60.99 & 12.13$\pm$0.46 & $-1.7$\,$\pm$0.12   \\
NGC 6715 & 8.29 & $-20.32$ & 11.25$\pm$0.59 & $-1.39$\,$\pm$ 0.10  \\  
Arp 2 & 12.38 & $-28.71$ & 11.96$\pm$0.51 & $-1.69$\,$\pm$ 0.14  \\
Whiting 1 & 168.23 & $-48.97$ & 6.5$\pm$0.75 & $-0.68$\,$\pm$0.03  \\
Terzan 8 & 8.83 & $-34.94$ & 12.89$\pm$0.43 & $-2.18$\,$\pm$ 0.23 \\
  \hline   
  \end{tabular}
\end{table*}

\begin{figure*}
 \includegraphics[scale=0.8]{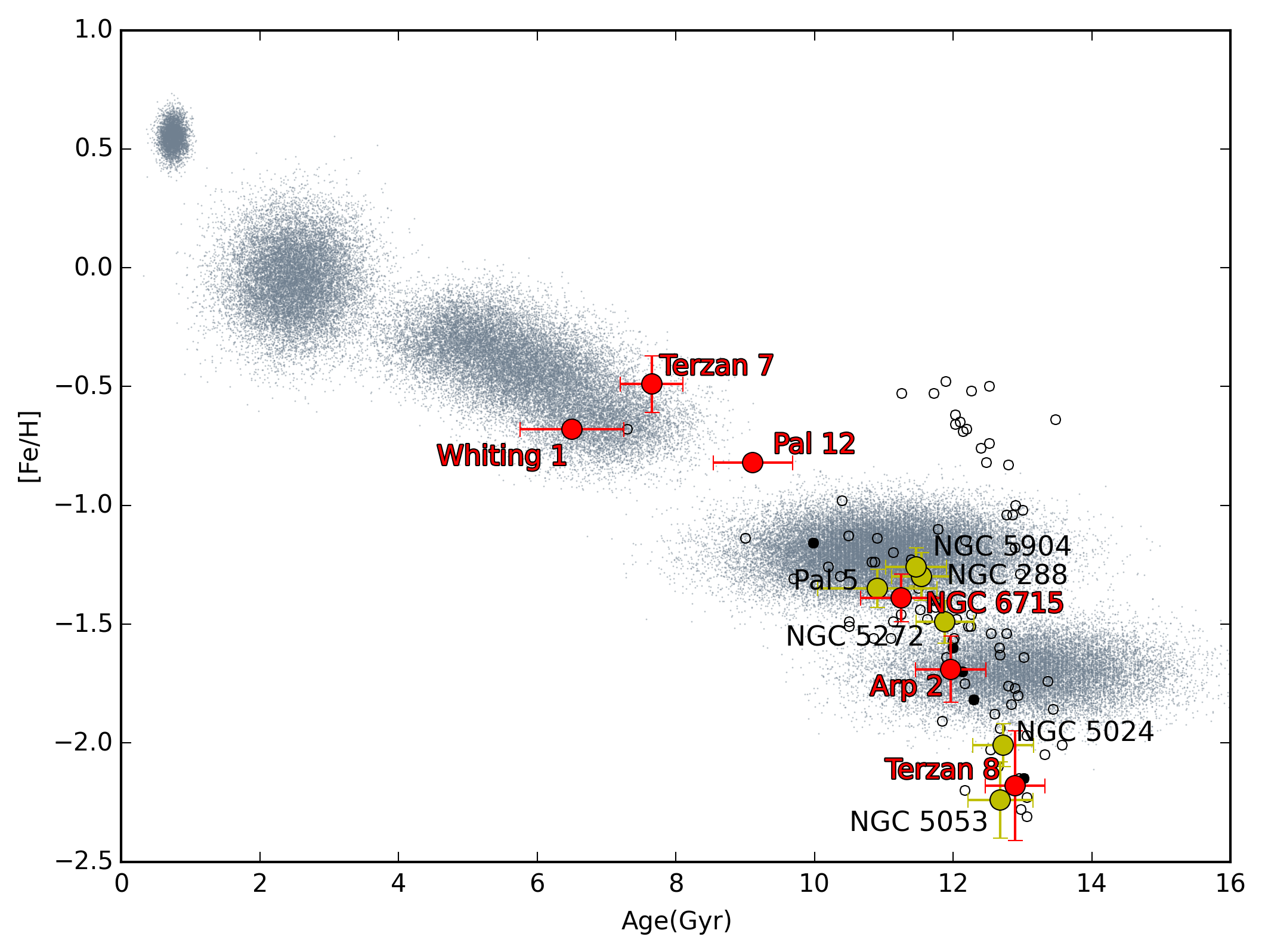} 
 \caption{ `` Age -- metallicity '' distribution. Gray dots represent the Sgr tidal stream model (LM10a). Red dots show six GCs from category A that belong to the stream, yellow dots show six clusters of category B, which differ in radial velocities or in three -- dimensional velocities. Black dots (five GCs) show clusters of category C, which differ in kinematics. Empty black circles are the remaining GCs of the Galaxy.}
 \label{fig:4}
\end{figure*}

We consider `` age -- metallicity '' distribution (see Fig.~\ref{fig:4}) for GCs \citep{2019MNRAS.486.3180K} and particles in the LM10a model as another  way to test the hypothesis that the GCs belong to the Sagittarius stream. Data on age \citep{2010MNRAS.404.1203F} and  metallicity  for 17 GCs can be seen in Table~\ref{tab:3}.   The average metallicity for all GCs is taken from the  catalog of \cite{1996AJ....112.1487H} (2010 Version) (http://physwww.mcmaster.ca/~harris/Databases.html) and the  papers  of \cite{2010MNRAS.404.1203F,2010ApJ...708..698D,2011ApJ...738...74D,2013ApJ...775..134V}. The separate groups of model stars in Fig.~\ref{fig:4} represent different star formation episodes. The position in this figure of GCs, close in coordinates, radial velocities and proper motions to Sgr dSph (see Sec.~\ref{sec:2.2}), approximately coincides with the position of groups of model stars, that is, starbursts. We see that the ages for all GCs clearly coincide with the star formation episodes. For several GCs, we obtain metallicity, which is significantly different from metallicity for particles in the model. Namely, their metallicity is slightly lower than for the model. This fact may mean that their formation was somewhat ahead of the formation of the main mass of old stars. However, these discrepancies are not very large, if we take into account the measurement errors of metallicity. For one GC out of six GCs in category A, which belong to the stream by kinematics, we obtain metallicity somewhat lower than for the model; this is $Terzan$~8   with metallicity $[Fe/H]=-2.18$~dex. For one of the clusters of category B, which includes six objects (see Sec.~\ref{sec:2.2}), a deviation from the general ``age -- metallicity'' relation is observed.  This is $NGC$~5053   with metallicity $[Fe/H]=-2.24$~dex. For one of the clusters of category C, which includes five objects (see Sec.~\ref{sec:2.2}), a deviation from the general ``age -- metallicity'' dependency is also observed for $ NGC $~5466 with metallicity of $[Fe/H]=-2.15$~dex.

We can also notice in Fig.~\ref{fig:4} that among the candidates we found, there are no GCs younger than approximately 6~Gyr. This may mean that either  clusters  did  not  form  during  this  period, or they were destroyed, or changed their trajectories as a result of interaction with the dense gas layers of the Galactic disk (Section 7 and Fig.11 in \cite{2010ApJ...718.1128L}  and \cite{2018MNRAS.478.5263T} article).  Note that the stage of the closest tidal interaction of Sgr dSph with our Galaxy was 
about 3~Gyr ago \citep[see for e.g.][]{2017ApJ...847...42D}.


\begin{figure*}
 \includegraphics[width=0.65\textwidth, angle=270]{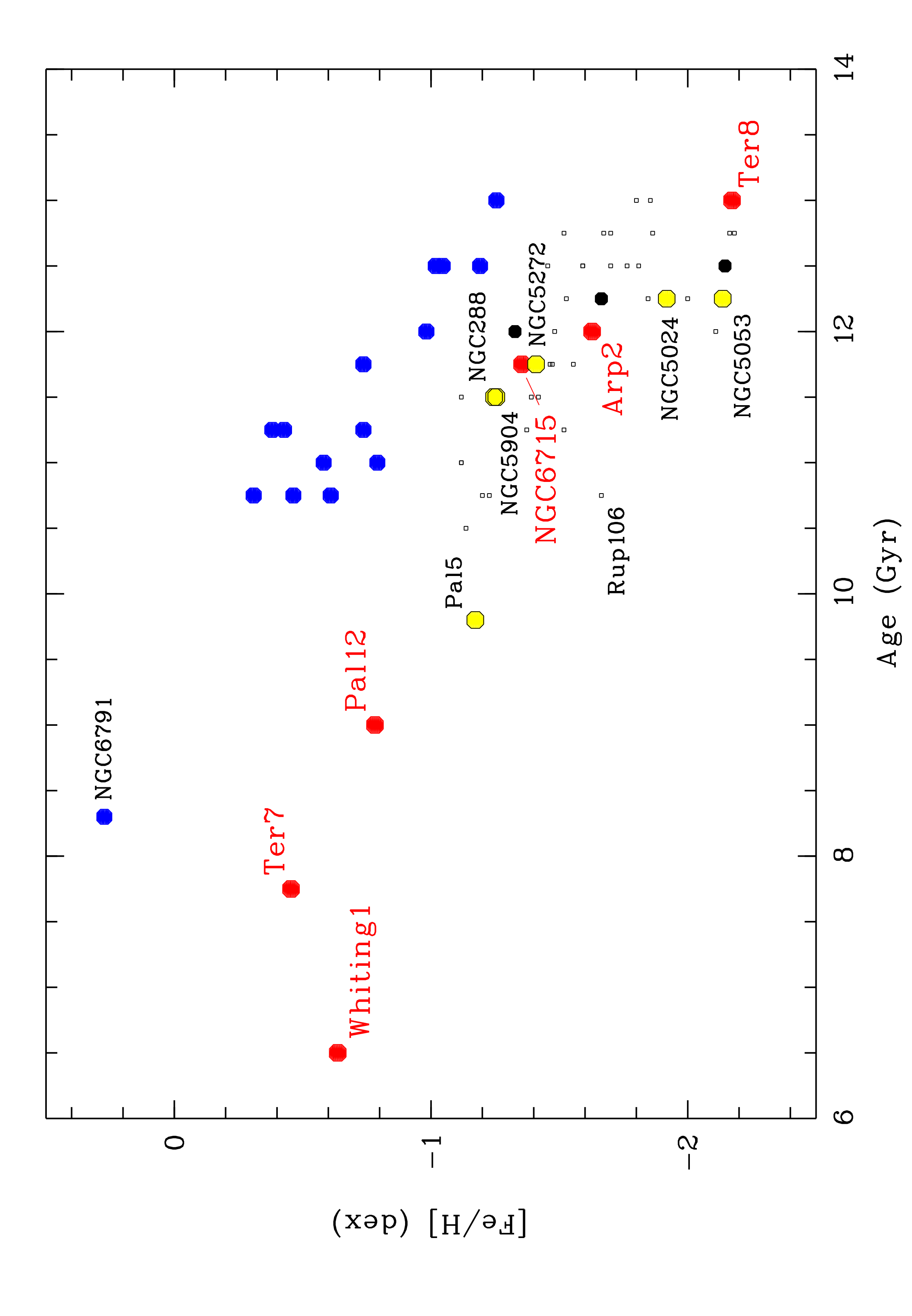}
 \caption{ `` Age -- metallicity ''  distribution according to data of \cite{2013ApJ...775..134V} and \cite{2013MNRAS.436..122L}. The red, black, yellow and empty black  symbols are the same clusters as in Fig. 4. Blue dots show high -- metallicity clusters belonging to the disk  according to \cite{2013ApJ...775..134V} and \cite{ 2013MNRAS.436..122L}  }
 \label{fig:5}
\end{figure*}

The division of the GCs of the Galaxy into approximately two groups has long been discussed in the literature: (1) GCs following the AMR of Sgr dSph,  Canis Major and other probable accreted satellites of our Galaxy;  (2) GCs having an age comparable to the age of the Universe, with different metallicities \citep[see for e.g.][]{2010MNRAS.404.1203F}. The first group is associated with dwarf satellites that fell on the Galaxy, the largest of which, apparently, were Sgr dSph and Canis Major.  The second group formed \textit{in situ}. 
Fig.~\ref{fig:5} reproduces the `` age -- metallicity '' relation of Milky Way GCs according to data of \cite{2013ApJ...775..134V} and \cite{2013MNRAS.436..122L}.  Data for $Whiting$~1 are taken from \cite{2007A&A...466..181C}. Estimates of age and metallicity are made in a single system. The authors of estimations note that the relation is divided into two: with one branch running from approximately 12.5 Gyr at $[Fe/H]=-1.7$~dex to 11 Gyr at $[Fe/H]=-1.2$~dex, while the other is offset to higher metallicities by about 0.6~dex at a fixed age. Blue icons indicate clusters which with high probability belong to the disk by their kinematics, i.e. formed \textit{in situ} \citep{2013MNRAS.436..122L}.  $NGC$~6791 is an old open cluster. Red, black, and yellow symbols in Fig.~\ref{fig:5} show the same clusters as in Fig.~\ref{fig:4}. The age and metallicity of $Terzan$~8 and $M$~30 are approximately the same according to \cite{2013ApJ...775..134V}. We also noted on Fig.~\ref{fig:5} a low -- mass Galactic halo, for example $Rup$~106. It was accreted rather than formed \textit{in situ} \citep{2010MNRAS.404.1203F}. Objects with the lowest metallicity  at a given age (left sequence) belong to the Galactic halo population, most likely consisting of clusters of accreted satellites. It can be seen in this figure, that all the GCs selected in this paper as the members of Sgr dSph lie on the left sequence. Some of them, for example, $Whiting$~1, are younger in age than other objects of the left sequence with a given metallicity.

High -- resolution spectroscopy showed \citep{2017A&A...608A.145B,2018ApJ...859L..10C,2018ApJ...855...83H,2019ApJ...872...58H},  that Sgr dSph stars have chemical compositions that are significantly different from those of stars in the Galaxy. For example, the content of $ \alpha $ -- process elements ($Mg$, $Si$, $Ca$, $Ti$) is much lower than that of Galactic stars with a given metallicity \citep[see for e.g.][]{2018ApJ...859L..10C,2018ApJ...855...83H,2019ApJ...872...58H}.  We demonstrate this effect for GCs in Sgr dSph using the example of the dependency of the titanium content $[Ti/Fe]$ from metallicity $[Fe/H]$. We chose the $Ti$ chemical element because the uncertainties in the estimation contents as a function of temperature, $\log g$, and metallicity are lower for $Ti$, than for other $ \alpha $ -- elements \citep{2014A&A...562A..71B}.  While significant anticorrelation in light element abundances are found in almost all of the old massive GCs of the Galaxy, the contents of the heaviest $ \alpha $ -- elements ($Si$, $Ca$, $Ti$), iron peak elements and other heavier elements vary slightly in the stars of these objects \citep[see for e.g.][]{2018ARA&A..56...83B}.

\begin{figure*}
 \includegraphics[scale=0.29,angle=270]{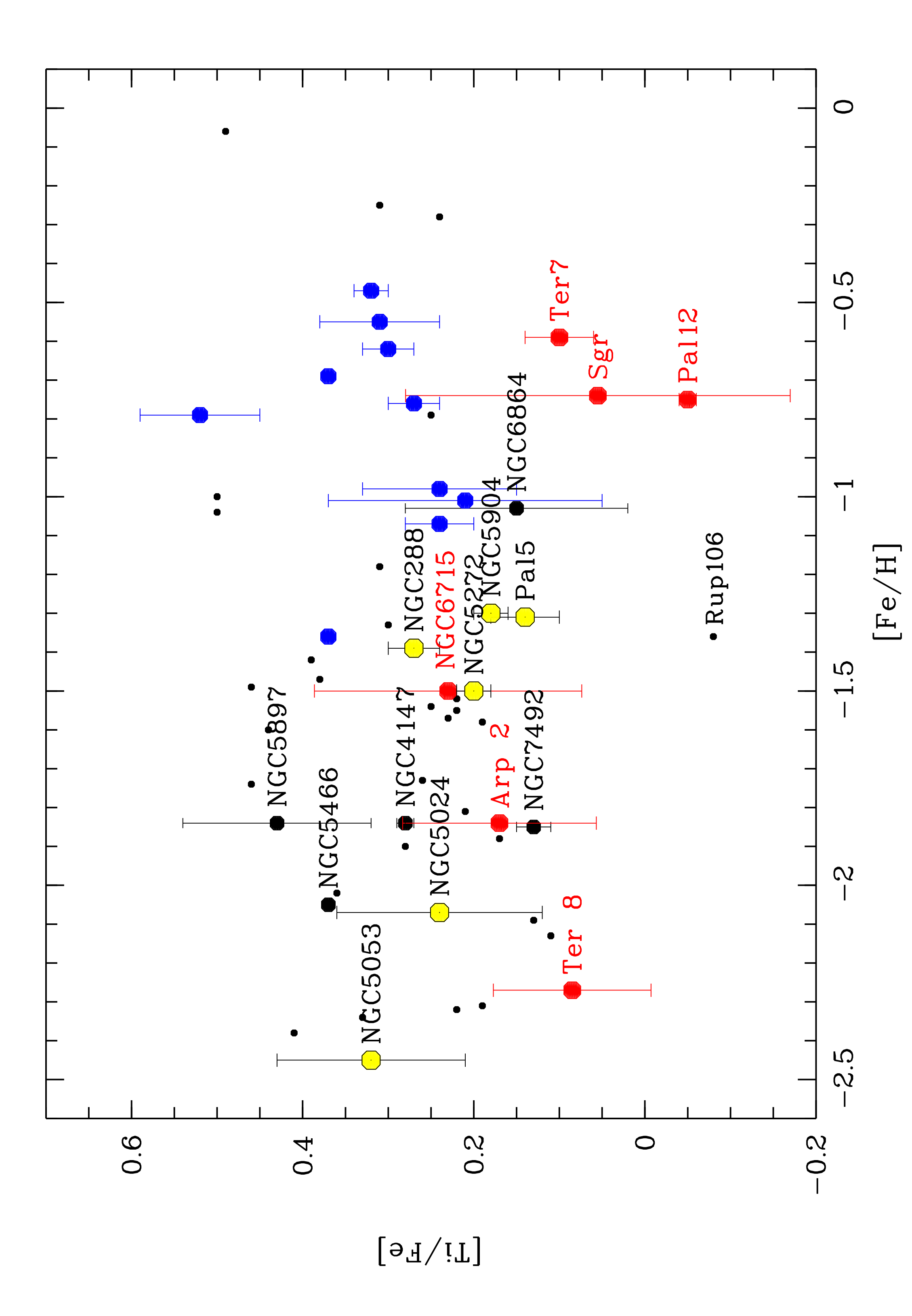} 
 \includegraphics[scale=0.29,angle=270]{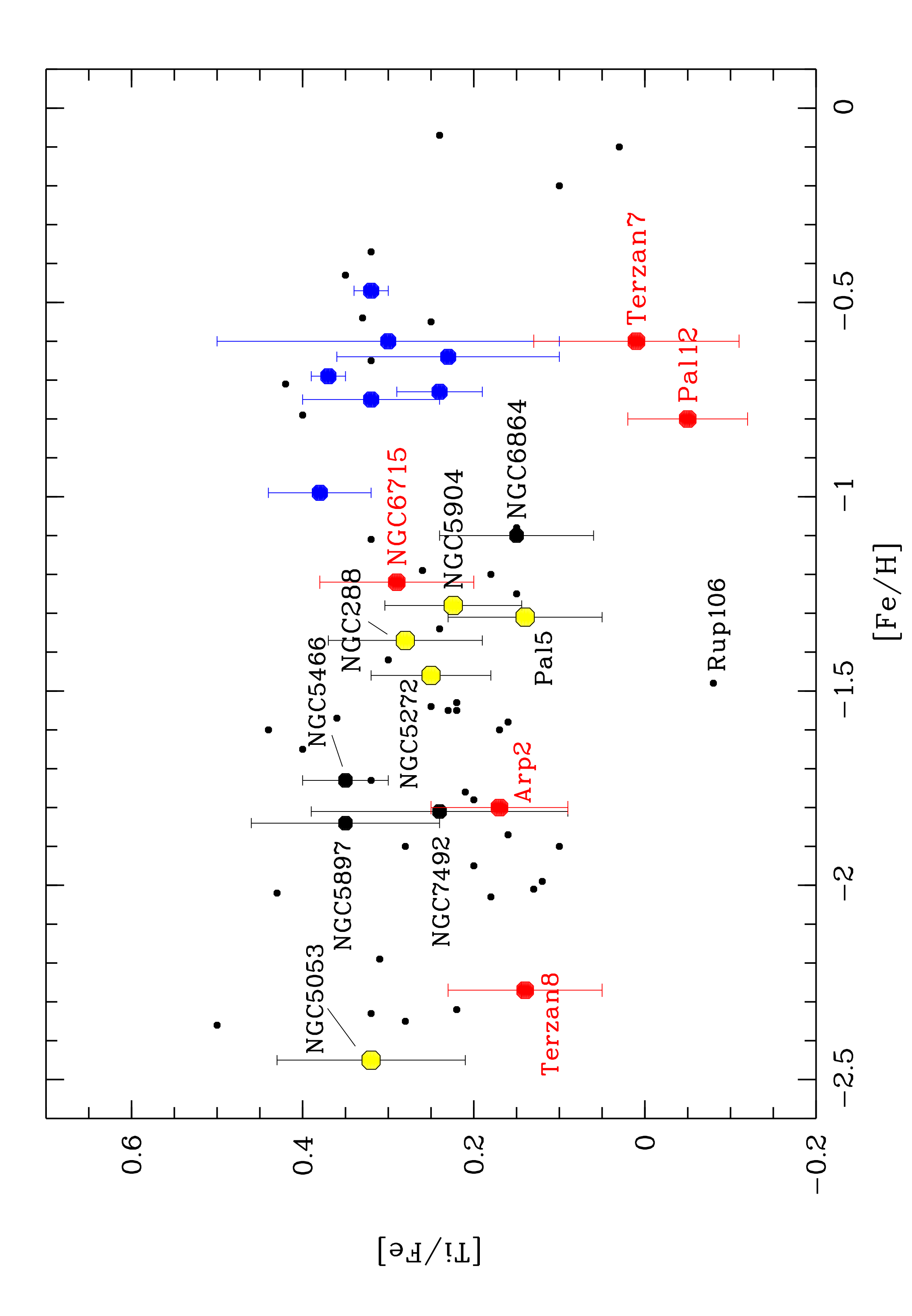} 
 \caption{ The dependency of [Ti/Fe] on [Fe/H] according to data collected by us from different sources (Fig. 6a, left figure) and according to data from the compilation catalog \cite{2019ARep...63..274M} (Fig. 6 b, right figure). The red, black, yellow and empty black  symbols are the same objects as in Fig. 4. Blue dots show high -- metallicity clusters belonging to the disk  according to \cite{2013ApJ...775..134V} and \cite{2013MNRAS.436..122L}  }
 \label{fig:6}
\end{figure*}

 The data of $[Ti/Fe]$ are mainly taken from \cite{2005AJ....130.2140P}. The average contents of $Ti$  are taken from the following papers:

-- for Sgr dSph, $Terzan$~8,  $Arp$~2, $NGC$~6715 from \cite{2014A&A...561A..87C},

-- for $NGC$~5927 from \cite{2018MNRAS.474.4541M},

-- for $NGC$~6362 from \cite{2017MNRAS.468.1249M} and \cite{2005AJ....130.2140P},

-- for $NGC$~6366 from \cite{2016AJ....152...21J} and \cite{2018MNRAS.476..690P},

-- for $NGC$~6624 from \cite{2011MNRAS.414.2690V},

-- for $NGC$~6717 from  \cite{1988AJ.....96.1925C},

-- for $NGC$~6723 from \cite{2016A&A...587A..95R},

-- for $NGC$~5024 from \cite{2015AJ....149..153M} and \cite{2016ApJ...824....5B},

-- for $NGC$~4147 from \cite{2016MNRAS.460.2351V},

-- for $NGC$~6791 from \cite{2007A&A...473..129C},

-- for $NGC$~4590 from \cite{2015AJ....149..204S},

-- for $NGC$~6528 from \cite{2018A&A...620A..96M},

-- for $NGC$~6864 from \cite{2013A&A...554A..81K}.

Figures~\ref{fig:6}~a and \ref{fig:6}~b show the dependence of $Ti$ contents on metallicity for GCs of the Galaxy according to data collected by us from different sources (Fig.~\ref{fig:6}~a) and according to the data from the compilation catalog \cite{2019ARep...63..274M} (Fig.~\ref{fig:6}~b). \cite{2019ARep...63..274M} contains no data for the following objects: $NGC$~4147, $NGC$~5024 and $NGC$~6723. Despite the large errors in the contents, on the whole, the dependences of $[Ti/Fe]$ on $[Fe/H]$ presented in the two panels of Fig.~\ref{fig:6} are similar. $[Ti/Fe]$ values in clusters, kinematically and spatially with high probability belonging to Sgr dSph (red symbols), are on average sistematically lower than for high -- metallicity clusters belonging to the disk  according to \cite{2013ApJ...775..134V} and \cite{2013MNRAS.436..122L} (blue dots in Fig.~\ref{fig:5} and \ref{fig:6}).  Some less likely members of stream (yellow and black large circles in Fig.~\ref{fig:6}) also have low $[Ti/Fe]$.

 Summarizing, it should be noted that the literature data on the ages, metallicities and contents of light elements in GCs, selected in this paper as the members of Sgr dSph, in total confirm the choice made.

\subsection{Results}\label{sec:2.4} 
As a result, we can conclude that we get three categories of GCs:

\textit{A: most certainly in the stream}, six GCs: $Terzan$~8; $Whiting$~1; $Arp$~2; $NGC$~6715; $Terzan$~7  and  $Pal$~12. The selected clusters coincide in all the parameters: by spatial positions, by position on the `` age -- metallicity '' dependencies, by radial velocities and by their proper motions. With a high probability, they belong to the Sgr tidal stream. For these clusters, we can surely call $Sgr$ $dSph$ as the host galaxy.

\textit{B: kinematic outliers}, six GCs: $Pal$~5; $NGC$~5904;   $NGC$~5024;  $NGC$~5053;  $NGC$~5272 and $NGC$~288. These are GCs that coincide in spatial positions, in position on the  `` age -- metallicity '' dependencies, but differ  in proper motions. So with low probability they belong to the stream and they can be referred to the list of candidates.

\textit{C: lowest rank candidates}, five GCs: $NGC$~6864; $NGC$~5466; $NGC$~5897; $NGC$~7492;  and $NGC$~4147. These clusters coincide in position on the `` age -- metallicity '' dependencies, and the probabilities of spatial positions in the stream are high (for example, the probability of spatial position in the stream is higher for $ NGC $~4147 than for $ Terzan $~7, which is exactly in the stream), but they diverge in radial and 3D velocities. Thus, these five lower -- ranking candidates with a high probability did not previously belong to the tidal stream of Sgr.

\section{Discussion and Conclusion}\label{sec:3}  
We conducted a search  for GCs belonging to the tidal stream of Sgr, which are currently scattered throughout the Milky Way. For this, we studied the GCs system in our Galaxy (157 GCs) and the Sgr stream, for which there are real data on stars (202 stars in the arms) and LM10a model data ($10^5$ particles). The GCs that came to our Galaxy with the Sgr dSph may still conserve the memory of their past host in their spatial distribution and kinematics. To identify the GCs belonging to the Sgr stream, we obtained 17 GCs using the method of nearest neighbors  and spatial density information for stars from the LM10a model. After that, for 17 GCs, the spatial distributions,  radial velocity distributions, proper motions, and position on the `` age -- metallicity '' dependencies in relation to the same parameters of stars from real dataand  for model LM10a were analyzed. As a result, we get three categories of GCs (A, B, C), six clusters in the first category, six in the second and five in the third, A -- the best candidates, and C -- the worst.

Our list of GCs belonging to the Sgr tidal stream (category A) is in good agreement with the lists obtained in the \cite{2010ApJ...718.1128L,2014MNRAS.445.2971C,2019A&A...630L...4M,2010MNRAS.404.1203F,2020MNRAS.tmp..233F,2020A&A...636A.107B} works  by several other methods.  \cite{2010MNRAS.404.1203F}  studied the age -- metallicity dependency and investigated the horizontal branch morphology.  \cite{2010ApJ...718.1128L} used dynamic models in combination with three--dimensional position and velocity data for GCs of the Galaxy and dSph galaxies to identify the Galaxy satellites that had originally formed in the Sgr dSph gravitational potential well and  were stripped from Sgr. They also studied the  `` age -- metallicity '' dependancy.   \cite{2014MNRAS.445.2971C} presented wide--field photometry for 23 GCs.  \cite{2019A&A...630L...4M} combined the kinematic information provided by Gaia, studied the age -- metallicity dependency and analyzed the dynamic properties of GCs. \cite{2020MNRAS.tmp..233F} used  the integrals of motion, the age -- metallicity dependency and $ \alpha $ -- element ratios.   \cite{2020A&A...636A.107B}, used RR Lyrae variables to trace the stream in 6D, and selected clusters that correspond to the observed stream in position and velocity.

All GCs from our category A are assigned to the stream in these works. However, there are several clusters, which are most often classified in the literature as belonging to the Sgr stream, but not getting into category A. These clusters are 
$ NGC $~4147 \cite{2010MNRAS.404.1203F,2004AJ....127.3394F,2005MNRAS.360..631M,2014MNRAS.445.2971C,2003AJ....125..188B}, $ NGC $~5634 \cite{2010MNRAS.404.1203F,2014MNRAS.445.2971C,2010ApJ...718.1128L,2003AJ....125..188B} and $ NGC $~2419 \cite{2019AstBu..74..403M,2018ApJ...862...52S,2017A&A...598L...9M,2014MNRAS.437..116B,2003ApJ...596L.191N,2020MNRAS.tmp..233F}. 

GC $ NGC $~4147, although  it  belongs  to  the  stream according to position on the sky and according to the ``age -- metallicity''  relation,  has  discrepancies  in  the average radial velocity of the nearest six observed stars and   in   its   spatial   velocity.  Our conclusion that $ NGC $~4147 does not belong to Sgr is consistent with the result of \cite{2020arXiv200111564R}.  We did not consider GC $ NGC $~5634, since it had a low probability (0.009) of belonging to the stream in spatial position. According to kinematics, it turns out that its radial velocity coincides with the average radial velocity of the nearest six stars in the stream within $ 3\sigma $ with the error estimation method we adopted, but does not coincide in 3D velocities.

As for the cluster $ NGC $~2419, we did not  considerit, since it has a very small probability of belonging to the stream in spatial position  and it has no nearest neighbors -- the observed stars from the stream. In the article \cite{2017A&A...598L...9M} $ NGC $~2419 is assigned to the stream based on the similarity of the orbits of the Sgr galaxy and this GC, although it is said that $ NGC $~2419 is at a much greater distance than the current Sgr orbit.  This is possible if Sgr was much more massive in the past and its debris would then occupy a large range of distances. Evidence in favor of this assumption is discussed by \cite{2019AstBu..74..403M,2017A&A...605A..46M,2017MNRAS.464..794G}, but our method does not use this assumption, since in the LM10 model the initial mass of Sgr dSph is small,  $6\times10^8 M_\odot$ .

The article \cite{2018MNRAS.481..918A} presents the spatial distribution of GCs in the Galaxy using the analysis of the inertia tensor, where it is shown that at a distance of less than $18$~kpc there is a statistically significant anisotropy, the distribution is flattened to the disk. In addition, at a distance of about 18~kpc, a feature is observed in the distribution of GCs: the ratio of the axes of the eigenvalues of the inertia tensor sharply changes, the  reason  for  which  is  unclear.  We suggest that this may be due to the passage of tidal streams at this distance, in particular to the passage of the tidal stream of Sgr. Further study of the effect of Sgr GCs on the spatial distribution of  GCs in the Galaxy will shed light on this issue.

\section{Funding}
The work was financially supported of the RAS project KP 19-270 ``Questions of the origin and evolution of the Universe using methods of ground observations and space research''.
M.E. Sharina thanks the RFBR for the grant 18--02--00167a.

\onecolumngrid
\clearpage

\bibliographystyle{AstroBull}
\bibliography{gc.bib}

\label{lastpage}
\end{document}